\documentstyle[prd,aps,epsfig]{revtex}
\newcommand{\prepr}[1] {\begin{flushright}  {\bf #1} \end{flushright}
\vskip 1.cm}
\newcommand{\titul}[1] {\begin{center}{\Large {\bf #1 } } \end{center}
\vskip 0.8cm}

\newcommand{\autor}[1] {\begin{center}  {\bf \lineskip .3cm #1  }
                        \end{center} }

\newcommand{\lugar}[1] {\begin{center}  {\normalsize \bf \it #1   }
\end{center}}
\begin{document}
\hbadness=10000
\pagenumbering{arabic}
\begin{titlepage}
\prepr{
\hspace{20mm} DPNU-00-13}
\titul{\bf
$B \to D^{(*)}$ form factors in perturbative QCD}
\autor{T. Kurimoto$^1$\footnote{E-mail: krmt@sci.toyama-u.ac.jp},
Hsiang-nan Li$^{2}$ \footnote{E-mail: hnli@phys.sinica.edu.tw} and
A.I. Sanda$^3$\footnote{E-mail: sanda@eken.phys.nagoya-u.ac.jp}
}
\lugar{ $^{1}$ Department of Physics, Toyama University, Toyama 930-8555,
Japan}
\lugar{ $^{2}$ Institute of Physics, Academia Sinica,
Taipei, Taiwan 115, Republic of China}
\lugar{ $^{2}$ Department of Physics, National Cheng-Kung University,\\
Tainan, Taiwan 701, Republic of China}
\lugar{ $^{3}$ Department of Physics, Nagoya University, Nagoya 464-8602,
Japan}

\vskip 2.0cm
{\bf  PACS index : 13.25.Hw, 11.10.Hi, 12.38.Bx, 13.25.Ft}

\thispagestyle{empty}
\vspace{10mm}
\begin{abstract}

We calculate the $B\to D^{(*)}$ form factors in the heavy-quark and
large-recoil limits in the perturbative QCD framework based on
$k_T$ factorization theorem, assuming the hierachy 
$M_B\gg M_{D^{(*)}}\gg \bar\Lambda$, with the $B$ meson mass $M_B$, the 
$D^{(*)}$ meson mass $M_{D^{(*)}}$, and the heavy meson and heavy quark
mass difference $\bar\Lambda$. The qualitative behavior of the light-cone
$D^{(*)}$ meson wave function and the associated Sudakov resummation are
derived. The leading-power contributions to the $B\to D^{(*)}$ 
form factors, characterized by the scale 
$\bar\Lambda\sqrt{M_B/M_{D^{(*)}}}$, respect the heavy-quark symmetry. 
The next-to-leading-power corrections in $1/M_B$ and $1/M_{D^{(*)}}$, 
characterized by a scale larger than $\sqrt{\bar\Lambda M_B}$, are 
estimated to be less than 20\%. The $D^{(*)}$ meson wave function is
determined from the fit to the observed $B\to D^{(*)} l\nu$ decay 
spectrum, which can be employed to make predictions for nonleptonic 
decays, such as $B\to D^{(*)}\pi(\rho)$.

\end{abstract}
\thispagestyle{empty}
\end{titlepage}


\section{INTRODUCTION}

Recently, we have made theoretical progress in the perturbative QCD 
(PQCD) approach to heavy-to-light decays, such as the proof of $k_T$ 
factorization theorem \cite{NL}, the construction of power counting rules
for various topologies of decay amplitudes \cite{CKL}, the derivation of
$k_T$ and threshold resummations \cite{LY1,TLS,L3}, and the
application to heavy baryon decays \cite{SLL}
and to three-body nonleptonic decays \cite{CL2}. 
Important dynamics in the decays $B\to K\pi$  and $B\to \pi\pi$ 
has been explored, including penguin enhancement and large CP 
asymmetries \cite{CKL,KLS,LUY,Keum}. It is worthwhile to investigate to 
what extent this formalism can be generalized to charmful decays,
such as the semileptonic decay $B\to D^{(*)}l\nu$ and the nonleptonic
decay $B\to D^{(*)}\pi(\rho)$. $k_T$ factorization theorem for the 
$B\to D^{(*)}$ form factors in the large-recoil region of the $D^{(*)}$ 
meson can be proved following the procedure in \cite{NL}, which 
are expressed as the convolution of hard amplitudes with $B$ and 
$D^{(*)}$ meson wave functions. A hard amplitude, being infrared-finite, 
is calculable in perturbation theory. The $B$ and $D^{(*)}$ meson wave 
functions, collecting the infrared divergences in the decays, are 
not calculable but universal.

The $B\to D^{(*)}$ transitions are more complicated than the 
$B\to\pi$ ones, because they involve three scales: the $B$ meson mass 
$M_B$, the $D^{(*)}$ meson mass $M_{D^{(*)}}$, and the heavy meson and 
heavy quark mass difference $\bar\Lambda=M_B-m_b\sim M_{D^{(*)}}-m_c$,
where $m_b$ ($m_c$) is the $b$ ($c$) quark mass, and $\bar\Lambda$ of
order of the QCD scale $\Lambda_{\rm QCD}$. There are several
interesting topics in the large-recoil region of the $B\to D^{(*)}$
transitions. (1) How to construct reasonable power counting rules for 
these decays with the three scales? (2) What is
the qualitative behavior of the wave function for an energetic
$D^{(*)}$ meson?  Is it dominated by soft dynamics the same as a $B$
meson wave function, or by collinear dynamics the same as a pion
wave function? (3) How different are the end-point 
singularities in the $B\to D^{(*)}$ form factors from those in the 
$B\to\pi$ ones? (4) what is the effect of Sudakov ($k_T$ and
threshold) resummation in the $B\to D^{(*)}$ form factors? (5) How are the 
relations for the $B\to D^{(*)}$ form factors in the heavy-quark limit
modified by the finite $B$ and $D^{(*)}$ meson masses? (6) What is the hard
scale for the $B\to D^{(*)}$ transitions? Is it large enough to make
sense out of perturbative evaluation of hard amplitudes? These questions
will be answered in this work.

We attempt to develop PQCD formalism for the 
$B\to D^{(*)}l\nu$ decays in the heavy-quark and large-recoil limits 
based on $k_T$ factorization theorem \cite{BS,LS}. We argue that the 
following hierachy must be postulated:
\begin{eqnarray}
M_B\gg M_{D^{(*)}}\gg \bar\Lambda\;.
\label{ll}
\end{eqnarray}
The relation $M_B\gg M_{D^{(*)}}$ justifies the perturbative analysis 
of the $B\to D^{(*)}$ form factors at large recoil and the definition of
light-cone $D^{(*)}$ meson wave functions. The relation 
$M_{D^{(*)}}\gg\bar\Lambda$ justifies the power expansion in the 
parameter $\bar\Lambda/M_{D^{(*)}}$. We shall calculate the
$B\to D^{(*)}$ form factors as double expansions in $M_{D^{(*)}}/M_B$
and in $\bar\Lambda/M_{D^{(*)}}$. The small ratio $\bar\Lambda/M_B
=(M_{D^{(*)}}/M_B)(\bar\Lambda/M_{D^{(*)}})$ is regarded as being
of higher power. Whether the hierachy is reasonable can be examined by
the convergence of high-order corrections in the strong coupling
constant $\alpha_s$ and of higher-power corrections in 
$1/M_B$ and in $1/M_{D^{(*)}}$. Note that Eq.~(\ref{ll}) differs from 
the small-velocity limit considered in \cite{SV}, where the ratio 
$M_{D^{(*)}}/M_B$ is treated as being of $O(1)$.

Under the hierachy, the wave function for an energetic $D^{(*)}$ meson
absorbs collinear dynamics, but with the $c$ quark line being 
eikonalized. That is, its definition is a mixture of those for a
$B$ meson dominated by soft dynamics and for a pion dominated by 
collinear dynamics. We examine the behavior of the heavy meson wave 
functions in the heavy-quark and large-recoil limits. For 
$\bar\Lambda/M_B$, $\bar\Lambda/M_{D^{(*)}}\ll 1$, only a single $B$ meson
wave function $\phi_B(x)$ and a single $D^{(*)}$ meson wave function 
$\phi_{D^{(*)}}(x)$ are involved in the $B\to D^{(*)}$ form factors, $x$
being the momentum fraction associated with the light spectator quark. 
Equations of motion for the relevant nonlocal matrix elements imply that 
$\phi_B(x)$ and $\phi_{D^{(*)}}(x)$ exhibit maxima at 
$x\sim \bar\Lambda/M_B$ and at $x\sim \bar\Lambda/M_{D^{(*)}}$, respectively. 
Since PQCD is reliable in the large-recoil region, it is powerful 
for the study of two-body nonleptonic $B$ meson decays. After determining
the $D^{(*)}$ meson wave function (the $B$ meson and pion wave functions
have beed fixed in our previous works), we can make predictions for the
$B\to D^{(*)}\pi(\rho)$ decays. Preliminary results for the $B\to D\pi$
modes will be presented in Sec.~IV.

Similar to the heavy-to-light transitions, the $B\to D^{(*)}$ form
factors also suffer singularities from the end-point region with a
momentum fraction $x\to 0$ in collinear factorization theorem. When the
end-point region is important, the parton transverse momenta $k_T$ are
not negligible, and $k_T$ factorization is a tool more appropriate
than collinear factorization. It has been proved that 
predictions for a physical quantity derived from $k_T$ factorization are 
gauge-invariant \cite{NL}. Because of the inclusion of parton transverse
degrees of freedom, the large double logarithmic corrections
$\alpha_s\ln^2 k_T$ appear and should be summed to all orders. It
turns out that the resultant Sudakov factor for an energetic $D^{(*)}$ 
meson is similar to that for a $B$ meson. Including the Sudakov effects 
from $k_T$ resummation and from threshold resummation for hard amplitudes
\cite{L3,UL}, the end-point singularities do not exist, and soft
contributions can be suppressed effectively. 

We shall identify the leading-power and next-to-leading-power
(down by $1/M_B$ or by $1/M_{D^{(*)}}$) contributions to the 
$B\to D^{(*)}$ form factors, which are equivalent to the expansion in 
heavy quark effective theory \cite{IW,EH}. It will be shown that the 
leading-power factorization formulas satisfy the heavy-quark relations 
defined in terms of the Isgur-Wise (IW) function \cite{IW}. These
contributions are characterized by the scale 
$\bar\Lambda\sqrt{M_B/M_{D^{(*)}}}$, indicating that the applicability of
PQCD to the $B\to D^{(*)}$ form factors may be marginal. The 
next-to-leading-power corrections to the heavy-quark relations, 
characterized by a scale larger than $\sqrt{\bar\Lambda M_B}$,
can be evaluated more reliablly. These corrections amount only up to
20\% of the leading contribution, indicating that the power expansion in
$1/M_B$ and in $1/M_{D^{(*)}}$ works well. It should be stressed that
the above percentage is only indicative, since we have not yet been
able to explore the complete next-to-leading-power sources with the
current poor knowledge of nonperturbative inputs.


In Sec.~II we define kinematics and explain $k_T$ factorization theorem 
for the $B\to D^{(*)}l\nu$ decay. The power counting rules are
constructed. In Sec.~III we discuss the behavior of the $B$ and $D^{(*)}$
meson wave functions, and perform $k_T$ resummation associated with an
energetic $D^{(*)}$ meson. The leading-power and next-to-leading-power 
contributions to the $B\to D^{(*)}$ form factors are calculated in 
Sec.~IV. Section V is the conclusion.

\section{KINEMATICS AND FACTORIZATION}

We discuss kinematics of the $B\to D^{(*)}l\nu$ decay
in the large-recoil region. The $B$ meson momentum $P_1$ and the 
$D^{(*)}$ meson momentum $P_2$ are chosen, in the light-cone
coordinates, as
\begin{eqnarray}
& &P_1=\frac{M_B}{\sqrt{2}}(1,1,0_T)\;,\;\;\;
P_2=\frac{r^{(*)}M_B}{\sqrt{2}}(\eta^+,\eta^-,0_T)\;,
\end{eqnarray}
with the ratio $r^{(*)}=M_{D^{(*)}}/M_B$. The factors 
$\eta^\pm=\eta\pm\sqrt{\eta^2-1}$ is defined in terms of the velocity
transfer $\eta=v_1\cdot v_2$ with $v_1=P_1/M_B$ and $v_2=P_2/M_{D^{(*)}}$.  
The longitudinal polarization vector $\epsilon_L$ and the transverse
polarization vectors $\epsilon_T$ of the $D^*$ meson are then given by
\begin{eqnarray}
\epsilon_L=\frac{1}{\sqrt{2}}(\eta^+,-\eta^-,0_T)\;,\;\;\;\;
\epsilon_T=(0,0,1)\;.
\end{eqnarray}

The partons involved in hadron wave functions are close to the mass 
shell \cite{CLY}. Assume that the heavy meson, the heavy quark, and the 
light spectator quark carry the momenta $P_H$, $P_Q=P_H-k$, and $k$, 
respectively. The on-shell conditions lead to
\begin{eqnarray}
& &k^2=O({\bar \Lambda}^2)\;,\;\;\;
\label{od0}\\
& &P_Q^2-m_Q^2=M_H^2-m_Q^2-2P_H\cdot k
=O({\bar\Lambda}^2)\;,
\label{od}
\end{eqnarray}
$M_H$ ($m_Q$) being the heavy meson (quark) mass. For the $B$ meson, we
have $2P_1\cdot k_1= 2 M_Bk_1^0\sim M_B^2-m_b^2=2M_B\bar\Lambda$ from 
Eq.~(\ref{od}), namely, $k_1^0\sim \bar\Lambda$.  The order of magnitude 
of the light spectator momentum is then, from Eq.~(\ref{od0}),
\begin{eqnarray}
k_1^\mu\sim (\bar\Lambda,\bar\Lambda,\bar\Lambda)\;.
\label{k1}
\end{eqnarray}
For the $D^{(*)}$ meson, we have 
$2P_2\cdot k_2= 2P_2^+k_2^- + 2 P_2^- k_2^+\sim 2 M_{D^{(*)}}
\bar\Lambda$, which implies
\begin{eqnarray}
k_2^\mu\sim \left(\frac{M_B}{M_{D^{(*)}}}\bar\Lambda,
\frac{M_{D^{(*)}}}{M_B}\bar\Lambda,\bar\Lambda\right)\;,
\label{k2}
\end{eqnarray}
for $\eta^+\sim 1/r^{(*)}$ and $\eta^-\sim r^{(*)}$.
With the above parton momenta, the exchanged gluon in the lowest-order 
diagrams is off-shell by
\begin{eqnarray}
(k_1-k_2)^2\sim -\frac{M_B}{M_{D^{(*)}}}\bar\Lambda^2\;,
\label{hs1}
\end{eqnarray}
which is identified as the characteristic scale of the hard amplitudes. 
To have a meaningful PQCD formalism, the large-recoil limit in 
Eq.~(\ref{ll}), $M_B\gg M_{D^{(*)}}$, is necessary.

We have proved $k_T$ factorization theorem for the semileptonic decay 
$B\to \pi l\nu$ \cite{NL}. Soft divergences from the region of a loop 
momentum $l$, where all its components are of $O(\bar\Lambda)$,
are absorbed into a light-cone $B$ meson wave function. Collinear
divergences from the region with $l$ parallel to the pion momentum in
the plus direction, whose components scale like
\begin{eqnarray}
l^\mu\sim (M_B,\bar\Lambda^2/M_B,\bar\Lambda)\;,
\label{sog2}
\end{eqnarray}
are absorbed into a pion wave function. The above meson 
wave functions, defined as nonlocal hadronic matrix elements, are 
gauge-invariant and universal. 

$k_T$ factorization theorem for the semileptonic decay
$B\to D^{(*)} l\nu$ is similar. In the limit $M_B\gg M_{D^{(*)}}$
we have $k_2^+\gg k_{2T}\gg k_2^-$ from Eq.~(\ref{k2}), indicating that
the $D^{(*)}$ meson wave function is dominated by collinear dynamics.
The leading infrared divergences in this decay are then classfied as
being soft, if a loop momentum $l$ vanishes like 
$l^\mu\sim (\bar\Lambda,\bar\Lambda,\bar\Lambda)$, and
as being collinear, if $l$ scales like $k_2$ in Eq.~(\ref{k2}). The
former (latter) are collected into a light-cone $B$ ($D^{(*)}$)
meson wave function. The collinear gluons defined by Eq.~(\ref{sog2}) 
do not lead to infrared divergences. 

Though the $D^{(*)}$ meson wave function absorbs
the collinear configuration, similar to the pion
wave function, the heavy-quark expansion applies to the $c$ quark
in the same way as to the $b$ quark in a $B$ meson. 
This is the reason we claim that the energetic $D^{(*)}$
meson dynamics is a mixture of those of the $B$ meson at rest and of
the energetic pion. Because $P_2\cdot l \sim M_{D^{(*)}}\bar \Lambda$ is 
much larger than $l^2\sim  \bar\Lambda^2$ according to Eq.~(\ref{ll}), 
we have the eikonal approximation,
\begin{eqnarray}
\frac{\not P_2-\not k_2 +\not l-m_c}{(P_2- k_2 + l)^2-m_c^2}\gamma^\alpha
c(P_2-k_2)\approx \frac{v_2^\alpha}{v_2\cdot l}c(P_2-k_2)\;,
\label{eik}
\end{eqnarray}
where $c(P_2-k_2)$ is the $c$ quark spinor, and the factor 
$v_2^\alpha/(v_2\cdot l)$ the Feynman rule for a rescaled $c$ quark
field. The physics involved in the above approximation is that the
kinematics of the spectator quark and of the $c$ quark is dramatically
different in the limit $M_{D^{(*)}}\to\infty$. Hence,  a gluon moving 
parallel to $k_2$ can not resolve the details of the $c$ quark, from
which its dynamics decouples.


According to $k_T$ factorization theorem \cite{NL}, the light spectator
momenta $k_1$ in the $B$ meson and $k_2$ in the $D^{(*)}$ meson are 
parametrized as
\begin{eqnarray}
k_1=\left(0,x_1\frac{M_B}{\sqrt{2}},{\bf k}_{1T}\right)\;,\;\;\;
k_2=\left(x_2\eta^+r^{(*)}\frac{M_B}{\sqrt{2}},0,{\bf k}_{2T}\right)\;,
\label{def}
\end{eqnarray}
where the momentum fractions $x_1$ and $x_2$ have the orders of magnitude,
\begin{eqnarray}
x_1\sim \bar\Lambda/M_B\;,\;\;\;\; x_2\sim \bar\Lambda/M_{D^{(*)}}\;.
\end{eqnarray}
The smallest component $k_2^-$ has been dropped. The neglect of $k_1^+$ 
is due to its absence in the hard amplitudes shown below.

\begin{figure}[t]
\centerline{
\includegraphics[width=10cm]{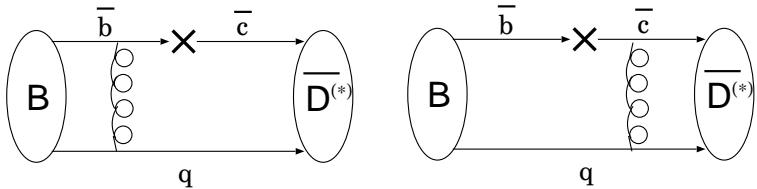}
}
\caption{Lowest-order diagrams for the $B\to D^{(*)}$ form factors.}
\label{fig1}
\end{figure}

The lowest-order diagrams for the $B\to D^{(*)}$ form factors are 
displayed in Fig.~1. The factorization formula is written as 
\begin{eqnarray}
\langle D^{(*)}(P_2)|{\bar b}\Gamma_\mu c|B(P_1)\rangle&=&
g^2C_F N_c\int dx_1 dx_2d^2k_{1\perp}d^2k_{2\perp}
\frac{dz^+d^2z_\perp}{(2\pi)^3}\frac{dy^-d^2y_\perp}{(2\pi)^3}
\nonumber \\
& & \times e^{-ik_2\cdot y}\langle D^{(*)}(P_2)|{\bar d}_{\gamma}(y)
c_{\beta}(0)|0\rangle
e^{ik_1\cdot z}\langle 0|{\bar b}_{\alpha}(0)d_{\delta}(z)|
B(P_1)\rangle~H_\mu^{\gamma\beta;\alpha\delta}\;,
\label{fbpi}
\end{eqnarray}
with the hard amplitude,
\begin{eqnarray}
H_\mu^{\beta\gamma;\delta\alpha}&=&[\gamma_\sigma]^{\gamma\delta}
\frac{1}{(k_2-k_1)^2}\left[\gamma^\sigma
\frac{\not k_2-\not P_1+m_b}{(P_1-k_2)^2-m_b^2}
\Gamma_\mu\right]^{\alpha\beta}
\nonumber\\
& &+[\gamma_\sigma]^{\gamma\delta}
\frac{1}{(k_2-k_1)^2}\left[\Gamma_\mu
\frac{\not k_1-\not P_2+m_c}{(P_2-k_1)^2-m_c^2}
\gamma^\sigma\right]^{\alpha\beta}\;,
\label{th}
\end{eqnarray}
for $\Gamma_\mu=\gamma_\mu$ or $\gamma_\mu\gamma_5$.
It is obvious that the large component $k_2^+$ picks only up the 
component $k_1^-$ in the denominators of the internal particle
propagators. The first and second terms in $H_\mu$ behave like 
$M_{D^{(*)}}^2/(\bar\Lambda^3 M_B^2)$ and $M_{D^{(*)}}/(\bar\Lambda^3M_B)$,
respectively. Therefore, for a leading-power formalism,
we keep $O(1)$ coefficients of the first term, and $O(r^{(*)})$ 
coefficients ($O(1)$ coefficients are absent) of the second term.

\section{HEAVY MESON WAVE FUNCTIONS}

In this section we discuss the qualitative behavior of the $B$, $D$ and 
$D^*$ meson wave functions in the heavy-quark and large-recoil limits, 
and derive $k_T$ resummation associated with the $D^{(*)}$ meson.

\subsection{$B$ Meson Wave Functions}

According to \cite{NL,GN,BF}, the two leading-twist $B$ meson wave
functions defined via the nonlocal matrix element,
\begin{eqnarray}
\int \frac{d^4w}{(2\pi)^4}e^{ik\cdot w}
\langle 0|{\bar b}_{\alpha}(0)d_{\delta}(w)|B^0(P)\rangle
&=&-\frac{i}{\sqrt{2N_c}}\left\{(\not P+M_B)\gamma_5
\left[\frac{\not{\bar n}}{\sqrt{2}}\phi_B^{+}(k)
+\frac{\not n}{\sqrt{2}}\phi_B^{-}(k)\right]
\right\}_{\alpha\delta}\;,
\label{bwp}\\
&=&\frac{i}{\sqrt{2N_c}}\{(\not P+M_B)\gamma_5
[\phi_B(k)+\not n{\bar \phi}_B(k)]\}_{\alpha\delta}\;,
\label{bwp1}
\end{eqnarray}
with the dimensionless vectors $\bar n=(1,0,0_T)$ and
$n=(0,1,0_T)$ on the light cone, and the wave functions, 
\begin{eqnarray}
\phi_B=\phi_B^{+}\;,\;\;\;
{\bar\phi}_B=(\phi_B^{-}-\phi_B^{+})/\sqrt{2}\;.
\end{eqnarray}
We have shown that the contribution from ${\bar\phi}_B$ starts from the
next-to-leading-power $\bar\Lambda/M_B$. This contribution, which may be 
numerically relevant \cite{WY}, should be included together with other
next-to-leading-power contributions in order to form a complete analysis. 
On this point, our opinion is contrary to that in \cite{DGS}.

The investigation based on equations of motion \cite{KQT} shows that
the distribution amplitude $\phi_B^{+}(x)=\int d^2k_T\phi_B^{+}(x,k_T)$ 
vanishes at the end points of the momentum fraction $x=k^-/P^-\to 0$, 1. 
Hence, we adopt the model in the impact parameter $b$ space
\cite{KLS},
\begin{eqnarray}
\phi_B(x,b)=N_Bx^2(1-x)^2
\exp\left[-\frac{1}{2}\left(\frac{xM_B}{\omega_B}\right)^2
-\frac{\omega_B^2 b^2}{2}\right]\;,
\label{os}
\end{eqnarray}
where the shape parameter $\omega_B$ has been determined as 
$\omega_B=0.4$ GeV. The normalization constant $N_B$ is related to 
the decay constant $f_B$ through
\begin{eqnarray}
\int dx\phi_B(x,b=0)=\frac{f_B}{2\sqrt{2N_c}}\;.
\end{eqnarray}
It is easy to find that $\phi_B$ in Eq.~(\ref{os}) has a maximum at
$x\sim \bar\Lambda/M_B$ as claimed in Sec.~II.

\subsection{$D$ Meson Wave Functions}

Consider the nonlocal matrix elements associated with the $D$ meson,
\begin{eqnarray}
\langle 0|{\bar c}(y)\gamma_5\gamma_\mu d(w)|D^-(P)\rangle&=&
-if_D P_{\mu}\int_0^1 dx
e^{-i (P-k)\cdot y-i k\cdot w}\phi_D^v(x)
\nonumber\\
& &-\frac{i}{2}f_D M_D^2\frac{z_\mu}{P\cdot z}
\int_0^1 dx e^{-i (P-k)\cdot y-i k\cdot w}g_D(x)\;,
\label{cv}\\
\langle 0|{\bar c}(y)\gamma_5 d(w)|D^-(P)\rangle&=&
-if_D m_{0}\int_0^1 dx
e^{-i (P-k)\cdot y-i k\cdot w}\phi_D^p(x)\;,
\label{bs}\\
\langle 0|{\bar c}(y)\gamma_5\sigma_{\mu\nu}d(w)|D^-(P)\rangle&=&
-\frac{i}{6}f_D m_{0}\left(1-\frac{M_D^2}{m_{0}^2}\right)
(P_{\mu}z_\nu-P_{\nu}z_\mu)
\nonumber\\
& &\times \int_0^1 dx
e^{-i (P-k)\cdot y-i k\cdot w}\phi_D^\sigma(x)\;,
\label{bt}
\end{eqnarray}
with $z=y-w$ and the $D$ meson decay constant $f_D$.
The light spectator $d$ quark carries the momentum $k$
with the momentum fraction $x=k^+/P^+$ and the $\bar c$ quark carries the
momentum $P-k$. In the heavy-quark limit we have
\begin{eqnarray}
m_0=\frac{M_D^2}{m_c+m_d}=M_D+O(\bar\Lambda)\;,
\end{eqnarray}
implying that the contribution from the distribution amplitude
$\phi_D^\sigma$ is suppressed by $O(\bar\Lambda/M_D)$ compared to those
from $\phi_D^v$ and $\phi_D^p$. The distribution amplitude $g_D$,
appearing at $O(r^2)$, is negligible. 

Rewrite the pseudo-tensor matrix element as
\begin{eqnarray}
\langle 0|{\bar c}(y)\gamma_5\sigma_{\mu\nu}d(w)|D^-(P)\rangle=
i\langle 0|{\bar c}(y)\gamma_5\gamma_\mu\gamma_\nu d(w)|D^-(P)\rangle
-ig_{\mu\nu}\langle 0|{\bar c}(y)\gamma_5 d(w)|D^-(P)\rangle\;,
\label{cb}
\end{eqnarray}
and differentiate both sides with respect to $w$ and $y$.
The differentiation on the left-hand side gives a result 
suppressed by $O(\bar\Lambda/M_D)$. The relations 
\begin{eqnarray}
& &\int dxx\phi_D^p(x)e^{-i (P-k)\cdot y-i k\cdot w}=O(\bar\Lambda/M_D)\;,
\label{ed1}\\
& &\int dx[\phi_D^p(x)-\phi_D^v(x)]e^{-i (P-k)\cdot y-i k\cdot w}
=O(\bar\Lambda/M_D)\;,
\label{ed3}
\end{eqnarray}
arise from the differentiation with respect to $w_\nu$ for $\mu=-$ 
and to $y_\mu$ for $\nu=+$, respectively. Equation (\ref{ed1}) states
that the distribution amplitude $\phi_D^p$ possesses a maximum at 
$x\sim \bar\Lambda/M_D$. Equation (\ref{ed3}) states that the moments 
of $\phi_D^p$ and $\phi_D^v$ differ by $O(\bar\Lambda/M_D)$ 
(they have the same normalizations). 

Neglecting the $O(\bar\Lambda/M_D)$ difference according to Eq.~(\ref{ll}),
only a single $D$ meson wave function is involved in the 
evaluation of the $B\to D$ form factors, 
\begin{eqnarray}
\int \frac{d^4w}{(2\pi)^4}e^{ik\cdot w}
\langle 0|{\bar c}_{\beta}(0)d_{\gamma}(w)|D^-(P)\rangle
=-\frac{i}{\sqrt{2N_c}}[(\not P+M_D)
\gamma_5]_{\gamma\beta}\phi_D(x)\;,
\end{eqnarray}
where the distribution amplitude,
\begin{eqnarray}
\phi_D=\frac{f_D}{2\sqrt{2N_c}}\phi_D^v=\frac{f_D}{2\sqrt{2N_c}}\phi_D^p\;,
\end{eqnarray}
satisfies the normalization,
\begin{eqnarray}
\int_0^1 dx\phi_D(x)=\frac{f_D}{2\sqrt{2N_c}}\;.
\end{eqnarray}
For the purpose of numerical estimate, we adopt the simple model,
\begin{eqnarray}
\phi_D(x)=\frac{3}{\sqrt{2N_c}}f_D x(1-x)[1+C_D(1-2x)]\;.
\label{phid}
\end{eqnarray}
The free shape parameter $C_D$ is expected to take a value, such that 
$\phi_D$ has a maximum at $x\sim \bar\Lambda/M_D\sim 0.3$.
We do not consider the intrinsic $b$ dependence of the $D$ meson wave 
function, which can be introduced along with more free
parameters. Note that Eq.~(\ref{phid}) differs from the
one of the Gaussian form proposed in \cite{LM}.

\subsection{$D^*$ Meson Wave Functions}

The information of the $D^*$ meson distribution amplitudes is extracted 
from equations of motions for the nonlocal matrix elements,
\begin{eqnarray}
\langle 0|{\bar c}(y)\gamma_\mu d(w)|D^{*-}(P,\epsilon)\rangle&=&
f_{D^*}M_{D^*}\left[P_{\mu}\frac{\epsilon\cdot z}{P\cdot z}
\int_0^1 dx e^{-i (P-k)\cdot y-ik\cdot w}\phi_{\parallel}(x)\right.
\nonumber\\
& &+\epsilon_{T\mu}\int_0^1 dx e^{-i (P-k)\cdot y-ik\cdot w}
g^{(v)}_\perp(x)
\nonumber\\
& &\left. -\frac{1}{2}z_\mu\frac{\epsilon\cdot z}{(P\cdot z)^2}
M^2_{D^*}\int_0^1 dx e^{-i (P-k)\cdot y-ik\cdot w}g_3(x)\right]
\label{dv}\\
\langle 0|{\bar c}(y)\sigma_{\mu\nu}d(w)|D^{*-}(P,\epsilon)\rangle&=&
if^T_{D^*}\left[(\epsilon_{T\mu}P_{\nu}-\epsilon_{T\nu}P_{\mu})
\int_0^1 dx e^{-i (P-k)\cdot y-ik\cdot w}\phi_\perp(x)\right.
\nonumber\\
& &+(P_{\mu} z_\nu-P_{\nu} z_\mu)\frac{\epsilon\cdot z}{(P\cdot z)^2}
M^2_{D^*}\int_0^1 dx e^{-i (P-k)\cdot y-ik\cdot w}h^{(t)}_{\parallel}(x)
\nonumber\\
& &\left. +\frac{1}{2}(\epsilon_{T\mu}z_\nu-\epsilon_{T\nu}z_{\mu})
\frac{M^2_{D^*}}{P\cdot z}\int_0^1 dx
e^{-i (P-k)\cdot y-ik\cdot w}h_3(x)\right]\;,
\label{dt}\\
\langle 0|{\bar c}(y)d(w)|D^{*-}(P,\epsilon)\rangle&=&
-\frac{i}{2}\left(f^T_{D^*}-f_{D^*}\frac{m_c+m_d}{M_{D^*}}\right)
\epsilon\cdot zM^2_{D^*}\int_0^1 dx
e^{-i (P-k)\cdot y-ik\cdot w} h^{(s)}_{\parallel}(x)\;,
\label{ds}\\
\langle 0|{\bar c}(y)\gamma_5\gamma_\mu d(w)|D^{*-}(P,\epsilon)\rangle
&=&-\frac{1}{4}\left(f_{D^*}-f_{D^*}^T\frac{m_c+m_d}{M_{D^*}}\right)
M_{D^*}\epsilon_{\mu}^{\nu\alpha\beta}\epsilon_{T\nu} P_{\alpha}
z_\beta\int_0^1 dx e^{-i (P-k)\cdot y-ik\cdot w}g^{(a)}_{T}(x)\;,
\label{da}
\end{eqnarray}
where the $D^*$ meson decay constant $f_{D^*}$ ($f_{D^*}^T$) is
associated with the longitudinal  (transverse) polarization.

In the heavy-quark limit we have
\begin{eqnarray}
f_{D^*}^T-f_{D^*}\frac{m_c+m_d}{M_{D^*}}\sim
f_{D^*}-f_{D^*}^T\frac{m_c+m_d}{M_{D^*}}\sim O(\bar\Lambda/M_{D^*})\;.
\end{eqnarray}
Hence, the contributions from the various distribution amplitudes are 
characterized by the powers,
\begin{eqnarray}
\phi_{\parallel}(x)\;, \;\;\;\phi_\perp(x)& :&\;\;\; O(1)\;,
\nonumber\\
g^{(v)}_\perp(x)\;, \;\;\;h^{(t)}_{\parallel}(x)& :&\;\;\; O(r^*)\;,
\nonumber\\
g_3(x)\;, \;\;\;h_3(x)& :&\;\;\; O(r^{*2})\;,
\nonumber\\
h^{(s)}_{\parallel}(x)\;, \;\;\;g^{(a)}_{T}(x)& :&\;\;\;
O(\bar\Lambda/M_B)\;.
\end{eqnarray}
To the current accuracy, we shall consider the distribution amplitudes
$\phi_{\parallel}(x)$ and $h^{(t)}_{\parallel}(x)$ for the longitudinal
polarization, and $\phi_\perp(x)$ and $g^{(v)}_\perp(x)$ for the
transverse polarization of the $D^*$ meson.

Rewrite the tensor matrix element as
\begin{eqnarray}
\langle 0|{\bar c}(y)\sigma_{\mu\nu}d(w)|D^{*-}(P)\rangle=
i\langle 0|{\bar c}(y)\gamma_\mu\gamma_\nu d(w)|D^{*-}(P)\rangle
-ig_{\mu\nu}\langle 0|{\bar c}(y)d(w)|D^{*-}(P)\rangle\;,
\label{cds}
\end{eqnarray}
and differentiate both sides with respect to $w$ and $y$.
The relations,
\begin{eqnarray}
& &\int dx xh_\parallel^{(t)}(x)e^{-i (P-k)\cdot y-ik\cdot w}
=O(\bar\Lambda/M_{D^*})\;,
\label{eds0}\\
& &\int dxx[\phi_\perp(x)+h_3(x)]e^{-i (P-k)\cdot y-ik\cdot w}
=O(\bar\Lambda/M_{D^*})\;,
\label{eds1}\\
& &\int dx[\phi_{\parallel}(x)-h_{\parallel}^{(t)}(x)]
e^{-i (P-k)\cdot y-ik\cdot w}=O(\bar\Lambda/M_{D^*})\;,
\label{eds2}\\
& &\int dx[\phi_{\parallel}(x)-g_3(x)]e^{-i (P-k)\cdot y-ik\cdot w}
=O(\bar\Lambda/M_{D^*})\;,
\label{eds3}\\
& &\int dx[\phi_\perp(x)+h_3(x)-2g_\perp^{(v)}(x)]
e^{-i (P-k)\cdot y-ik\cdot w}=O(\bar\Lambda/M_{D^*})\;,
\label{eds4}
\end{eqnarray}
come from the derivatives
with respect to $w_\nu$ for $\mu=-$, to $w_\nu$ for $\mu=\perp$, to 
$y_\mu$ for $\nu=-$, to $y_\mu$ for $\nu=+$, and to $y_\mu$ for 
$\nu=\perp$, respectively. Equations (\ref{eds0}) and (\ref{eds1}) 
indicate that the $D^*$ meson distribution amplitudes have maxima at 
$x\sim \bar\Lambda/M_{D^*}$. Equations (\ref{eds2}) and (\ref{eds3}) state
that $\phi_\parallel$, $h_\parallel^{(t)}$ and $g_3$ are identical up to 
corrections of $O(\bar\Lambda/M_{D^*})$. Similarly,
$\phi_\perp$, $h_3$ and $g_\perp^{(v)}$ are also identical up to 
corrections of $O(\bar\Lambda/M_{D^*})$ from Eq.~(\ref{eds4}). 

Neglecting the $O(\bar\Lambda/M_{D^*})$ difference, we consider the
structure for a $D^*$ meson,
\begin{eqnarray}
\int \frac{d^4w}{(2\pi)^4}e^{ik\cdot w}
\langle 0|{\bar c}_{\beta}(0)d_{\gamma}(w)|D^{*-}(P)\rangle
&=&-\frac{i}{\sqrt{2N_c}}[(\not P+M_{D^*})\not\epsilon_L\phi_{D^*}^L(x)
+(\not P+M_{D^*})\not\epsilon_T\phi_{D^*}^T(x)]_{\gamma\beta}\;,
\label{dss}
\end{eqnarray}
with the definitions,
\begin{eqnarray}
& &\phi_{D^*}^L=\frac{f_{D^*}}{2\sqrt{2N_c}}\phi_{\parallel}
=\frac{f_{D^*}^T}{2\sqrt{2N_c}}h_{\parallel}^{(t)}\;,
\\
& &\phi_{D^*}^T=\frac{f_{D^*}^T}{2\sqrt{2N_c}}\phi_\perp
=\frac{f_{D^*}}{2\sqrt{2N_c}}g_\perp^{(v)}\;.
\end{eqnarray}
The $D^*$ meson distribution amplitudes satisfy  the normalizations,
\begin{eqnarray}
\int_0^1 dx\phi_{D^*}^{L}(x)=\int_0^1 dx\phi_{D^*}^{T}(x)=
\frac{f_{D^*}}{2\sqrt{2N_c}}\;,
\end{eqnarray}
where we have assumed $f_{D^*}=f_{D^*}^T$. Note that equations of motion
do not relate $\phi_{D^*}^{L}$ and $\phi_{D^*}^{T}$.
In this work we shall simply adopt the same model,
\begin{eqnarray}
\phi_{D^*}^L(x)=\phi_{D^*}^T(x)=
\frac{3}{\sqrt{2N_c}}f_{D^*}x(1-x)[1+C_{D^*}(1-2x)]\;.
\label{phis}
\end{eqnarray}
Similarly, the free shape parameter $C_{D^*}$ is expected to take a 
value, such that  $\phi_{D^*}$ has a maximum at 
$x\sim \bar\Lambda/M_{D^*}\sim 0.3$.

\subsection{Sudakov Resummation}

Radiative corrections to the
meson wave functions and to the hard amplitudes generate double
logarithms from the overlap of collinear and soft enhancements. 
The double logarithmic corrections $\alpha_s\ln^2 k_T$ to the heavy and
light meson wave functions and their Sudakov resummation have been 
analyzed in \cite{LY1}. The property of the $D^{(*)}$ meson wave function
is special, since its dynamics is a mixture of the soft one in the $B$ 
meson wave function and the collinear one in the pion wave function. In 
this section we derive $k_T$ resummation for the $b$-dependent $D$ meson 
wave function,
\begin{eqnarray}
\phi_D(x,b)&=&\frac{i}{\sqrt{2N_c}}
\int\frac{dw^-}{2\pi}e^{ik_2^+w^-}
\langle 0|{\bar c}_{v_2}(0)\frac{\gamma_5\gamma^+}{2}
P\exp\left[ig\int_0^{w}ds\cdot A(s)\right]
d(w)|D^-(P_2)
\rangle\;,
\label{dw}
\end{eqnarray}
with the coordinate $w=(0,w^-,{\bf b})$. The path for the Wilson link is
composed of three pieces: from 0 to $\infty$ along the direction of
$n$, from $\infty$ to $\infty+{\bf b}$, and from
$\infty+{\bf b}$ back to $w$ along the direction of $-n$ \cite{NL}. The
derivation for the $D^{*}$ meson is the same. The
$D^{(*)}$ meson distribution amplitude $\phi_{D^{(*)}}(x)$ discussed
in the previous two subsections is regarded as the initial condition of
the Sudakov evolution in $b$. 

\begin{figure}[t]
\centerline{
\includegraphics[width=8cm]{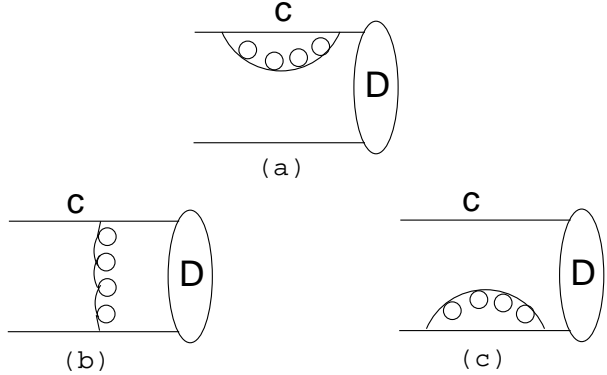}
}
\caption{Radiative corrections to the $D$ meson wave 
function in the axial gauge.}
\label{fig2}
\end{figure}

In the axial gauge the double logarithms appear in the two-particle 
reducible corrections, such as Figs.~2(b) and 2(c). Figure 2(a) gives 
only single soft logarithms. To implement the resummation technique, we 
allow the direction $n$ of the Wilson line to vary away from the light
cone. Because of the rescaled $\bar c$ quark field, $P_2$ does not lead to
a large scale, and the only large scale is $k_2$. Due to the scale
invariance in $n$ and in $v_2$, $k_2$ appears in the ratios
$(k_2\cdot n)^2/n^2$ and $(k_2\cdot v_2)^2/v_2^2$. However, the second 
ratio is of $O(r^2)$ compared to the first one, and negligible. Therefore, 
$\phi_D$ depends only on the single large scale $(k_1\cdot n)^2/n^2$. 
The rest part of the derivation then follows that in \cite{LY1}. The 
Sudakov factor from $k_T$ resummation is given by
\begin{eqnarray}
\exp[-S_{D^{(*)}}(\mu)]=\exp\left[-s(k_2^+,b)-2\int_{1/b}^\mu
\frac{d{\bar\mu}}{\bar\mu}\gamma(\alpha_s({\bar\mu}))\right]\;,
\label{ktd}
\end{eqnarray}
with the quark anomalous dimension $\gamma=-\alpha_s/\pi$. For
the explicit expression of the Sudakov exponent $s$, refer to \cite{KLS}.
It is found that Eq.~(\ref{ktd}) has the same functional form as the
Sudakov factor for the $B$ meson.

The double logarithms $\alpha_s\ln^2 x$ produced by the radiative 
corrections to the hard amplitudes are the same as in the 
$B\to\pi, \rho$ decays at leading power in $1/M_B$ and in $1/M_{D^{(*)}}$.
\cite{UL}. Threshold resummation of these logarithms leads to
\begin{eqnarray}
S_t(x)=\frac{2^{1+2c}\Gamma(3/2+c)}{\sqrt{\pi}\Gamma(1+c)}
[x(1-x)]^c\;,
\end{eqnarray}
with the constant $c=0.3\sim 0.4$. The factor $S_t(x_2)$ ($S_t(x_1)$),
associated with the first (second) term of $H_\mu$ in
Eq.~(\ref{th}), suppresses the end-point region with $x_2\to 0$
($x_1\to 0$). In the numerical study below we shall adopt $c=0.35$.

\section{$B\to D, D^*$ FORM FACTORS}

The $B\to D^{(*)}$ transitions are defined by the matrix elements,
\begin{eqnarray}
\langle D (P_2)|{\bar b}(0)\gamma_\mu c(0)|B(P_1)\rangle
&=&\sqrt{M_BM_D}\left[\xi_+(\eta)(v_1+v_2)_\mu+
\xi_-(\eta)(v_1-v_2)_\mu\right]\;,
\nonumber\\
\langle D^*(P_2,\epsilon^*)|{\bar b}(0)\gamma_\mu\gamma_5 c(0)|B(P_1)
\rangle&=&\sqrt{M_BM_{D^*}}
\left[{\xi_{A1}}(\eta)(\eta+1)\epsilon^*_\mu-{\xi_{A2}}(\eta)
\epsilon^*\cdot v_1v_{1\mu}-{\xi_{A3}}(\eta)\epsilon^*\cdot v_1 v_{2\mu}
\right]
\nonumber\\
\langle D^*(P_2,\epsilon^*)|{\bar b}(0)\gamma_\mu c(0)|B(P_1)\rangle&=&
i\sqrt{M_BM_{D^*}}\xi_V(\eta)\epsilon^{\mu\nu\alpha\beta}
\epsilon^*_\nu v_{2\alpha}v_{1\beta}\;.
\end{eqnarray}
The form factors $\xi_+$, $\xi_-$, $\xi_{A_1}$, $\xi_{A_2}$, $\xi_{A_3}$,
and $\xi_V$ satisfy the relations in the heavy-quark limit,
\begin{equation}
\xi_+=\xi_V=\xi_{A_1}=\xi_{A_3}=\xi,\;\;\;\;  \xi_-=\xi_{A_2}=0\;,
\label{iwr}
\end{equation}
where $\xi$ is the IW function \cite{IW}.

We write the form factors as the sum of the leading-power and
next-to-leading-power contributions,
\begin{equation}
\xi_i=\xi_i^{(0)}+\xi_i^{(1)}\;,
\label{iw01}
\end{equation}
for $i=+$, $-$, $A_1$, $A_2$, $A_3$, and $V$. 
The leading-power factorization formulas are given by
\begin{eqnarray}
\xi_+^{(0)}&=&16\pi C_F\sqrt{r}M_B^2\int dx_1dx_2
\int b_1db_1 b_2db_2\phi_B(x_1,b_1)\phi_D(x_2)
\nonumber\\
& &\times\left[E(t^{(1)}) h(x_1,x_2,b_1,b_2)
+rE(t^{(2)}) h(x_2,x_1,b_2,b_1)\right]\;,
\label{xip}\\
\xi_{-,A2}^{(0)}&=&0\;,
\label{xim}\\
\xi_{A1,A3,V}^{(0)}&=&16\pi C_F\sqrt{r^*}M_B^2\int dx_1dx_2
\int b_1db_1 b_2db_2\phi_B(x_1,b_1)\phi_{D^*}(x_2)
\nonumber\\
& &\times \left[E(t^{(1)})h(x_1,x_2,b_1,b_2)
+r^*E(t^{(2)})h(x_2,x_1,b_2,b_1)\right]\;,
\label{xia1}
\end{eqnarray}
with the color factor $C_F=4/3$. Obviously, the above expressions
obey the heavy-quark relations in Eq.~(\ref{iwr}), if we 
assume
\begin{eqnarray}
M_D=M_{D^*}\;,\;\;\;\;\phi_D=\phi_{D^*}\;.
\end{eqnarray}

The hard function is written as
\begin{eqnarray}
h(x_1,x_2,b_1,b_2)&=&K_{0}\left(\sqrt{x_1x_2r^{(*)}\eta^+}M_Bb_1\right)
S_t(x_2) 
\nonumber \\
& &\times \left[\theta(b_1-b_2)K_0\left(\sqrt{x_2r^{(*)}\eta^+}M_B
b_1\right)I_0\left(\sqrt{x_2r^{(*)}\eta^+}M_Bb_3\right)\right.
\nonumber \\
& &\left.+\theta(b_2-b_1)K_0\left(\sqrt{x_2r^{(*)}\eta^+}M_Bb_2\right)
I_0\left(\sqrt{x_2r^{(*)}\eta^+}M_Bb_1\right)\right]\;.
\label{dh}
\end{eqnarray}
In the evolution factor,
\begin{eqnarray}
E(t)=\alpha_s(t)\exp[-S_B(t)-S_{D^{(*)}}(t)]\;,
\end{eqnarray}
we keep the Sudakov factor associated with the $B$ meson, and allow the
behavior of the $B$ meson wave function to determine whether this effect 
is important. For the model $\phi_B(x,b)$ in
Eq.~(\ref{os}), the Sudakov effect is not important because of
$x\sim \bar\Lambda/M_B$. The hard scales $t$ are defined as
\begin{eqnarray}
t^{(1)}&=&{\rm max}(\sqrt{x_2r^{(*)}\eta^+}M_B,1/b_1,1/b_2)\;,
\nonumber\\
t^{(2)}&=&{\rm max}(\sqrt{x_1r^{(*)}\eta^+}M_B,1/b_1,1/b_2)\;.
\label{hat}
\end{eqnarray} 
Equations (\ref{xip}) and (\ref{xia1}) contain at most the logarithmic
end-point singularities in the collinear factorization, which are
weaker than the linear singularities in the $B\to\pi$ form factors. 
We conclude that Sudakov resummation is also crucial for the 
$B\to D^{(*)}$ transitions.

The next-to-leading-power corrections are given by
\begin{eqnarray}
\xi_+^{(1)}&=&4\pi C_F\sqrt{r}M_B^2\int dx_1dx_2\int b_1db_1 b_2db_2
\phi_B(x_1,b_1)\phi_D(x_2)
\nonumber\\
& &\times\frac{(\eta^+-1)(\eta^+-2)}{\sqrt{\eta^2-1}}
\left[rx_2E(t^{(1)}) h(x_1,x_2,b_1,b_2)
+x_1E(t^{(2)}) h(x_2,x_1,b_2,b_1)\right]\;,
\label{xip1}\\
\xi_-^{(1)}&=&-4\pi C_F\sqrt{r}M_B^2\int dx_1dx_2\int b_1db_1 b_2db_2
\phi_B(x_1,b_1)\phi_D(x_2)
\nonumber\\
& &\times\frac{(\eta^++1)(\eta^+-2)}{\sqrt{\eta^2-1}}
\left[rx_2E(t^{(1)}) h(x_1,x_2,b_1,b_2)
-x_1E(t^{(2)}) h(x_2,x_1,b_2,b_1)\right]\;,
\label{xim1}\\
\xi_{A1}^{(1)}&=&8\pi C_F\sqrt{r^*}M_B^2\int dx_1dx_2
\int b_1db_1 b_2db_2\phi_B(x_1,b_1)\phi_{D^*}(x_2)
\nonumber\\
& &\times \left[\frac{\eta^+-2}{\eta+1}r^*x_2
E(t^{(1)})h(x_1,x_2,b_1,b_2)
-\frac{x_1}{\eta+1}E(t^{(2)})h(x_2,x_1,b_2,b_1)\right]\;,
\label{xia11}\\
\xi_{A2}^{(1)}&=&-16\pi C_F\sqrt{r^*}M_B^2\int dx_1dx_2
\int b_1db_1 b_2db_2\phi_B(x_1,b_1)\phi_{D^*}(x_2)
\nonumber\\
& &\times \frac{\eta^+x_1}{\sqrt{\eta^2-1}}E(t^{(2)})h(x_2,x_1,b_2,b_1)\;,
\label{xia21}\\
\xi_{A3,V}^{(1)}&=&-8\pi C_F\sqrt{r^*}M_B^2\int dx_1dx_2
\int b_1db_1 b_2db_2\phi_B(x_1,b_1)\phi_{D^*}(x_2,b_2)
\nonumber\\
& &\times \left[\frac{\eta^+-2}{\sqrt{\eta^2-1}}r^*x_2
E(t^{(1)})h(x_1,x_2,b_1,b_2)
-\frac{x_1}{\sqrt{\eta^2-1}}E(t^{(2)})h(x_2,x_1,b_2,b_1)\right]\;.
\label{xia31}
\end{eqnarray}
whose hard amplitudes are consistent with those obtained in
\cite{KK,L1,WYL}. The additional powers in $x_2$ and in $x_1$ provide
stronger suppression in the end-point region with $x_1$, $x_2\to 0$.
Hence, the characteristic hard scales of the corresponding terms
increase to $\sqrt{\bar\Lambda M_B}$ and to
$M_B\sqrt{\bar\Lambda/M_{D^{(*)}}}$, respectively.

Consider the expansion of the currents to
$O(1/m_b)$ and $O(1/m_c)$ \cite{N1},
\begin{equation}
{\bar b}\Gamma_i c\approx
{\bar b}_{v_1}\Gamma c_{v_2}
+\frac{1}{2m_c}{\bar b}_{v_1}\Gamma_i i\not D_2 c_{v_2}
-\frac{1}{2m_b}{\bar b}_{v_1}\Gamma_i i\not D_1 c_{v_2}
+\cdots\;,
\label{hqet}
\end{equation}
where $\cdots$ stand for the terms, which are further suppressed
by $\alpha_s$. Compared to the right-hand side of Eq.~(\ref{hqet}),
$\xi_i^{(0)}$ and the terms proportional to $x_2$ ($x_1$) in 
$\xi_i^{(1)}$ are identified as the first term and the second (third) 
term. We do not distinguish $M_B$ and 
$m_b$, and $M_{D^{(*)}}$, $m_c$ and $m_0$, and employ only one 
single wave function for the $B$ and $D^{(*)}$ mesons. The differences 
of the above quantities are also the sources of $1/m$ corrections. 
However, their estimation requires more information of nonperturbative 
inputs, and can not be performed in this work. Equations 
(\ref{xip1})-(\ref{xia31}) will be 
employed to obtain an indication of the order of magnitude of  
next-to-leading-power corrections to the $B\to D^{(*)}$ form factors.

\begin{figure}[t]
\centerline{
\includegraphics[width=9cm]{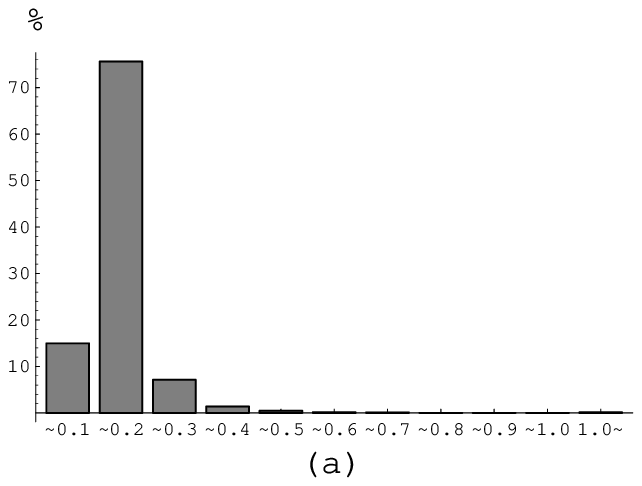}
\\
\includegraphics[width=9cm]{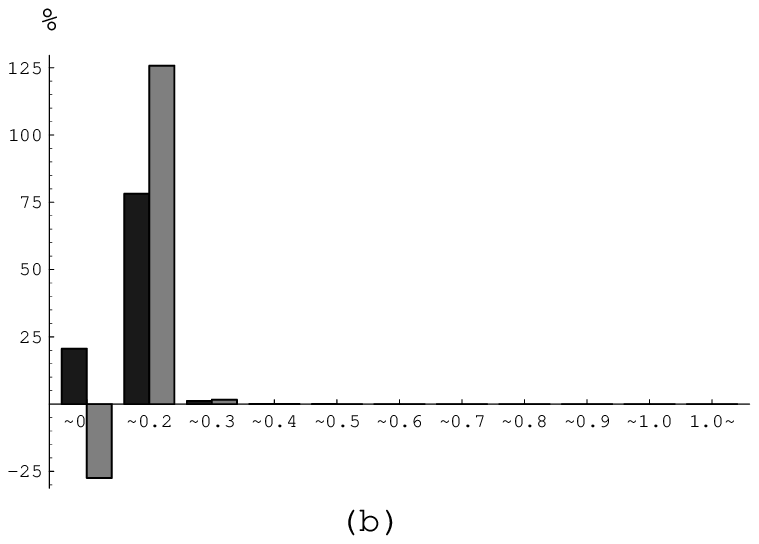}
}
\caption{(a) Contribution to $\xi_+^{(0)}$ at the maximal recoil from
the different ranges of $\alpha_s/\pi$.
(b) Contributions to $\xi_+^{(1)}$ (black) and to $\xi_-^{(1)}$ (gray)
 at the maximal recoil
from the different ranges of $\alpha_s/\pi$.}
\label{fig3}
\end{figure}

\begin{figure}[t]
\centerline{
\includegraphics[width=8cm]{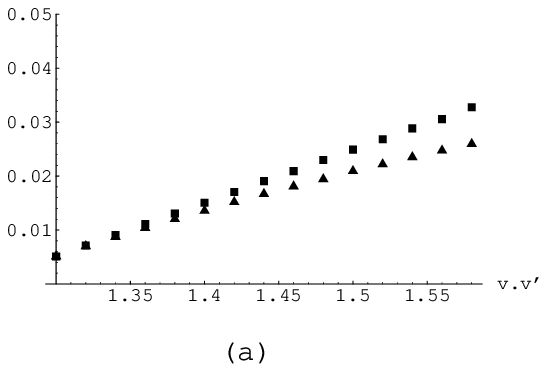}
\\
\includegraphics[width=8cm]{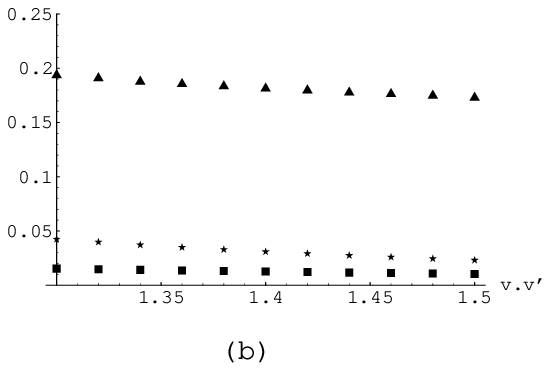}
}
\caption{(a) Dependence of $\xi_+^{(1)}/\xi_+^{(0)}$ (square) and of
$\xi_-^{(1)}/\xi_+^{(0)}$ (triangle) on the velocity transfer.
(b) Dependence of $-\xi_{A_1}^{(1)}/\xi_{A_1}^{(0)}$ (square), of
$-\xi_{A_2}^{(1)}/\xi_{A_1}^{(0)}$ (triangle), and of
$\xi_{A_3,V}^{(1)}/\xi_{A_1}^{(0)}$ (star) on the velocity transfer.}
\label{fig4}
\end{figure}

It is observed from Fig.~3(a) that most of the contribution to the
form factor $\xi_+^{(0)}$ comes from the range of $\alpha_s/\pi <0.3$,
implying that the applicability of PQCD to the $B\to D^{(*)}$ form 
factors is acceptable, and not worse than that to the $B\to\pi$ ones 
\cite{TLS}. This is attributed to the fact that the hard scales
$\bar\Lambda\sqrt{M_B/M_{D^{(*)}}}$ and $\sqrt{\bar\Lambda M_B}$ in the
two cases do not differ very much. The applicability improves for the
next-to-leading-power contributions as shown in Fig.~3(b): most of
them arise from $\alpha_s/\pi <0.2$. We emphasize that the above
percentage analysis is only indicative, and that the convergence of 
higher-order corrections needs to be justified by explicit calculation. 
We estimate from Fig.~4 that next-to-leading-power contributions are 
less than 20\% of the leading one. In fact, they are small except  
$\xi_{A2}^{(1)}$, implying that the power expansion in 
$\bar\Lambda/M_B$ and in $\bar\Lambda/M_{D^{(*)}}$ is reliable. 
Our results are smaller than those obtained
from QCD sum rules at large recoil \cite{BG}.

\begin{figure}[t]
\centerline{
\includegraphics[width=9cm]{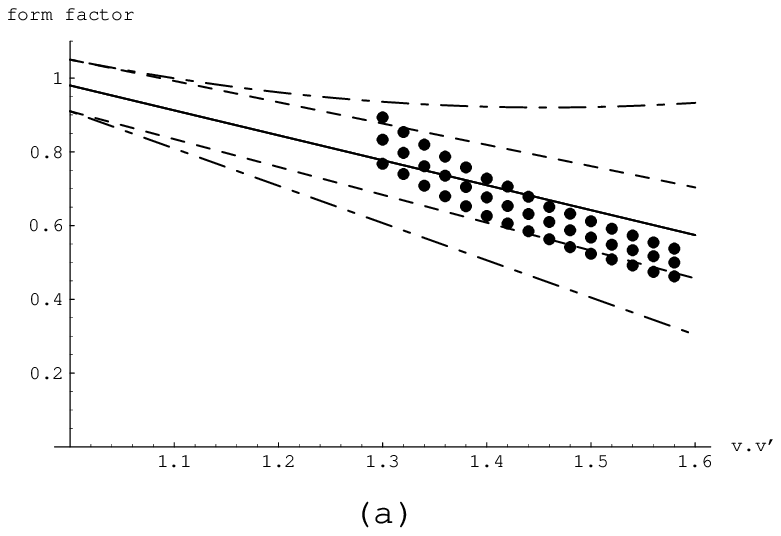}
\\
\includegraphics[width=9cm]{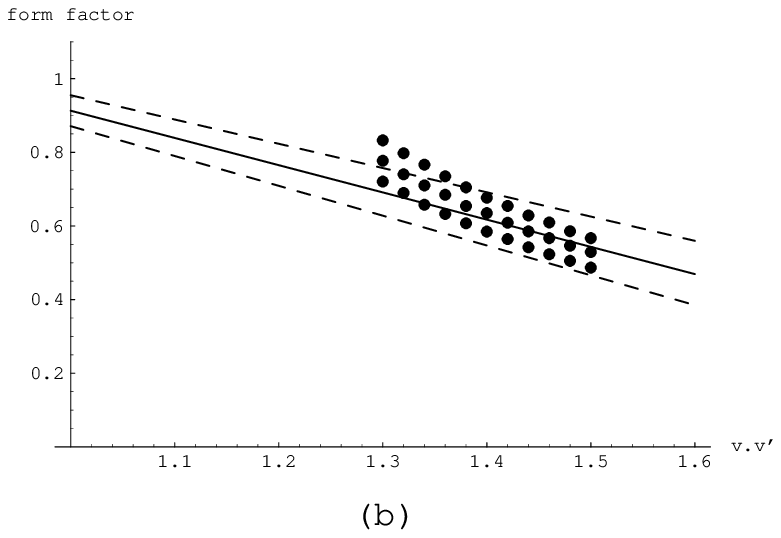}
}
\caption{(a) [(b)] $\xi$ as a function of the velocity transfer from the
$B\to D^{(*)}l\nu$ decay.
The solid lines represent the central values, the dashed
(dot-dashed) lines give the bounds from the linear (quadratic) fits.
The circles correspond to $C_{D^{(*)}}=0.5$, 0.7, and 0.9 from
bottom to top.}
\label{fig5}
\end{figure}

The next step is to determine the $D^{(*)}$ meson distribution amplitudes,
{\it i.e.}, the free parameters $C_{D^{(*)}}$, by fitting
the leading-power PQCD predictions to the measured decay spectra
at large recoil \cite{Beld,Belds,CLEO}. The
IW function extracted from the $B\to D^{(*)}l{\nu}$ decay is 
parametrized as 
\begin{eqnarray}
\xi(\eta)=F_{D^{(*)}}(1)[1-{\hat\rho}^2_{D^{(*)}}(\eta-1) 
+ {\hat c}_{D^{(*)}}(\eta-1)^2
+O((\eta-1)^3)]\;,
\end{eqnarray}
with the factors $F_{D}(1)=0.98\pm 0.07$ \cite{CLN} and 
$F_{D^*}(1)=0.913\pm 0.042$ \cite{NBa}. 
The above values are consistent with those derived from
lattice calculations \cite{SH1}.
The linear and quadratic fits give \cite{Beld,Belds}
\begin{eqnarray}
& &{\hat\rho}^2_{D}=0.69\pm 0.14\;,\;\;\;\;{\hat c}_{D}=0\;,\;\;\;\;
{\hat\rho}^2_{D^*}=0.81\pm 0.12\;,\;\;\;\;{\hat c}_{D^*}=0\;,
\nonumber\\
& &{\hat\rho}^2_{D}=0.69^{+0.42}_{-0.15}\;,\;\;\;\;
{\hat c}_{D}=0.00^{+0.59}_{-0.00}\;,
\end{eqnarray}
respectively. Choosing the decay constants $f_B=190$ MeV and 
$f_D=f_{D^*}=240$ MeV, we find that
$C_D\sim C_{D^*}=0.7$ leads to an excellent agreement with the data at
large recoil as exhibited in Fig.~5. For these values, the corresponding
$D^{(*)}$ meson distribution amplitude exhibits a maximum at $x\sim 0.36$,
consistent with our expectation. The rough equality of $C_D$ and
$C_{D^*}$ hints that the heavy-quark symmetry holds well.

Note that our aim is not to extract the Cabbibo-Kobayashi-Maskawa
matrix element $|V_{cb}|$ from experimental data. This extraction is
best done in the zero-recoil region, where the heavy-quark symmetry
defines unambiguously the normalization of the $B\to D^{(*)}$
transition form factors. It has been emphasized that the PQCD formalism 
is reliable at large recoil, and appropriate for two-body nonleptonic 
decays. Therefore, one of the purposes of this work is to determine the 
unknown $D^{(*)}$ meson wave function.
The $B$ meson and pion wave functions have been fixed already
in the literature. With these meson wave functions being available,
we are able to predict the branching ratios of
two-body nonleptonic charmful decays, such as $B\to D^{(*)}\pi(\rho)$.
Our predictions for the $B\to D\pi$ branching ratios \cite{L10},
\begin{eqnarray}
& &B(B^-\to D^{0}\pi^-)\sim 5.5\times 10^{-3}\;,
\nonumber\\
& &B({\bar B}^0\to D^{+}\pi^-)\sim 2.8\times 10^{-3}\;,
\nonumber\\
& &B({\bar B}^0\to D^{0}\pi^0)\sim 2.6\times 10^{-4}\;,
\label{pred}
\end{eqnarray}
are in agreement with experimental data \cite{BelleC,CLEOC,Bab}. The above
results correspond to the phenomenological coefficients $a_1$ and $a_2$ 
\cite{NPe} with the ratio $|a_2|/a_1\sim 0.5$ and the phase $-57^o$ of 
$a_2$ relative to $a_1$. The point is, from the view point of PQCD, that 
the phase is of short distance and generated from hard amplitudes. This is 
contrary to the conclusion drawn from naive factorization 
\cite{NPe,C02,X,CHY}: the phase comes from long-distance final-state 
interaction.

\begin{figure}[t]
\centerline{
\includegraphics[width=8cm]{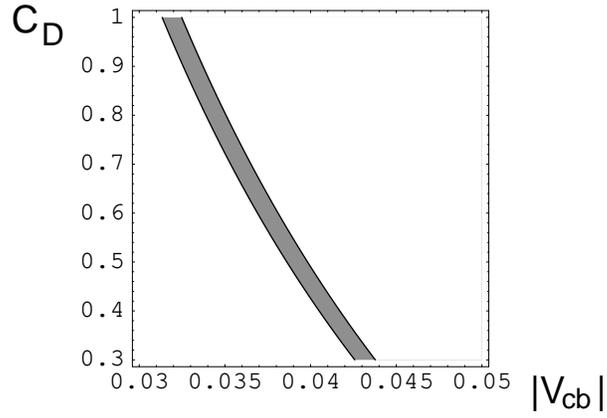}
}
\caption{The ${\tilde\chi}^2$ value as functions of
$|V_{cb}|$ and $C_D$ with the shaded
region corrsponding to ${\tilde\chi}^2 \le 1$.
}
\label{fig6}
\end{figure}

If the $D^{(*)}$ meson decay constant is known from, for example,
lattice QCD calculation, it is then possible to extract the matrix
element $|V_{cb}|$ from the measured semileptonic decay spectra at large
recoil using the PQCD formalism. The experimental data of the product
$[|V_{cb}|\xi(\eta)]_{\rm exp}$ for the $B\to Dl\nu$ mode \cite{Beld} 
are listed in Table~\ref{tab1}. The region with the large velocity 
transfer $\eta >1.35$ is regarded as the one, where PQCD analyses are 
reliable. We compute the following $\chi$-square as a function of the two 
parameters, $|V_{cb}|$ and the shape parameter $C_D$ ($N_{\rm para}=2$):
\begin{equation}
{\tilde\chi}^2 \equiv \frac{1}{ N_{\rm data} - N_{\rm para}}
\sum_{k} \left|\frac{{[|V_{cb}|\xi (\eta_k)}]_{\rm exp} - 
|V_{cb}| \xi_+^{(0)}(\eta_k, C_D)}{\sigma_k}\right|^2\;,
\end{equation}
where $N_{\rm data}=4$, $\sigma_k$ are assumed to be 10\% of the data 
(considering only the systematic errors for illustration), and
$\xi_+^{(0)}$ has been defined in Eq.~(\ref{xip}). The band 
in Fig.~\ref{fig6} stands for the
allowed $|V_{cb}|$-$C_D$ range with $\tilde \chi^2 \le 1$. It is found 
that taking $C_D=0.7$, $V_{cb}=(0.035\sim 0.0036)(240\;{\rm MeV}/f_D)$ is
consistent with the value extracted in \cite{Beld} from the zero-recoil
data. Choosing $V_{cb}=0.04$, $C_D=0.4\sim 0.5$ is allowed.

\begin{table}
\begin{center}
\begin{tabular}{c||cccc}
k&1&2&3&4\\\hline
$\eta$ &1.39&1.45&1.51&1.57\\
$[|V_{cb}|\xi (\eta)]_{\rm exp}$&0.028&0.030&0.024&0.024
\end{tabular}
\end{center} 
\caption{Experimantal data of $|V_{cb}|\xi(\eta)$ for the
$B\to Dl\nu$ decay.}
\label{tab1} 
\end{table} 


\section{CONCLUSION}

In this paper we have developed the PQCD formalism for the
$B\to D^{(*)}$ transitions in the heavy-quark and large-recoil limits based
on $k_T$ factorization theorem. The reasonable power counting rules for
these decays with the three scales $M_B$, $M_{D^{(*)}}$ and $\bar\Lambda$
have been constructed following the hierachy in Eq.~(\ref{ll}). Under this
hierachy, only a single $B$ meson wave function and a single $D^{(*)}$
meson wave function are involved, which possess maxima at the spectator 
momentum fractions $x\sim \bar\Lambda/M_B$ and
$x\sim \bar\Lambda/M_{D^{(*)}}$, respectively. Dynamics
of an energetic $D^{(*)}$ meson is the mixture of those of a $B$ meson
at rest and of an energetic light meson: it absorbs collinear divergences
but the heavy-quark expansion applies to the $c$ quark. The Sudakov 
factor from $k_T$ resummation for an energetic 
$D^{(*)}$ meson is similar to that associated with a $B$ meson.
The end-point singularities, being logarithmic in collinear 
factorization theorem, do not exist in $k_T$ factorization theorem.
Including also the Sudakov effect from threshold resummation for hard
amplitudes, the PQCD approach to the $B\to D^{(*)}$ transitions
becomes more reliable.

The factorization formulas for the $B\to D^{(*)}$ form factors have been 
expressed as the sum of leading-power and next-to-leading-power
contributions, which is equivalent to the heavy-quark expansion in both
$1/m_b$ and $1/m_c$. The leading-power formulas, respecting
the heavy-quark symmetry, are identified as the IW function. This
contribution, characterized by the scale
$\bar\Lambda\sqrt{M_B/M_{D^{(*)}}}$, is calculable marginally in PQCD. 
The next-to-leading-power corrections, characterized by a scale larger 
than $\sqrt{\bar\Lambda M_B}$, can be estimated more reliably, and found 
to be less than 20\% of the leading contribution. That is,
the heavy-quark expansion makes sense. Note that the
next-to-leading-power corrections considered here, which can be analyzed 
under the current knowledge of nonperturbative inputs, are not complete. 
The conclusion drawn in this paper
provides a solid theoretical base for the PQCD analysis of
the $\Lambda_b$ baryon charmful decays \cite{SLL2}.

We have determined the $D^{(*)}$ meson wave function from the
$B\to D^{(*)}l\nu$ decay spectrum, which has
a maximum at the spectator momentum fraction $x\sim 0.36$ as expected.
This wave function is useful for making predictions for the two-body
nonleptonic decays in the PQCD formalism. The results of the $B\to D\pi$
branching ratios have been presented in Eq.~(\ref{pred}), which are
consistent with the experimental data.
The detail of this subject will be published elsewhere.

\vskip 0.5cm
\centerline{\large\bf Acknowledgements}

The work of H.N.L. was supported in part by the National Science Council
of R.O.C. under the Grant No. NSC-91-2112-M-001-053, by National
Center for Theoretical Sciences of R.O.C., and by Theory Group of KEK, 
Japan. The work of T.K was supported in part by Grant-in Aid for 
Scientific Research from the Ministry of Education, Science and Culture 
of Japan under the Grant No. 11640265.

\end{document}